\documentstyle[12pt,epsf]{article}
\begin{document}
\input epsf
\title{Abelian Chern-Simons term in Superfluid $^{3}$He-A.}
\author{J. Goryo\thanks{e-mail:goryo@particle.sci.hokudai.ac.jp}
~and~K. Ishikawa\thanks{e-mail:ishikawa@particle.sci.hokudai.ac.jp} \\
{\it Department of Physics, Hokkaido University,}\\
{\it Sapporo, 060 Japan}}
\maketitle

\begin{abstract}

 We show in this paper that Abelian Chern-Simons term is induced 
in 2+1 and 3+1 dimensional rotating superfluid $^{3}$He-A and plays 
important roles in its dynamics.  
Because U(1) symmetry is spontaneously broken in $^{3}$He-A, 
Goldstone mode appears and contributes to induced Chern-Simons term. 
We found that the coefficient of Chern-Simons term, which is equivalent to 
Hall conductance, depends on an infra-red cut off 
of the Goldstone mode, and that the orbital angular momentum of $^{3}$He-A 
in a cylinder geometry is derived from Chern-Simons term. 

\end{abstract}

\section{Introduction} 

Abelian Chern-Simons term is induced in 2+1 dimensional parity violating 
electron system, and is the lowest dimensional gauge invariant 
object\cite{2+1C-S}. 
Its roles has been studied in anyon superconductivity\cite{anyon}
and quantum Hall effect\cite{QHE}.
In the quantum Hall system an effect 
of Chern-Simons term has been observed. One can expect that 
there may be other physical systems in which Chern-Simons term plays 
some roles.

It is known that 
Chern-Simons term has non-trivial topological structure, 
and the coefficient of induced term becomes topological invariant 
and not affected from higher order corrections in the gauge invariant systems,
without spontaneous symmetry breaking. In fact, these assumptions 
are realized in quantum Hall system and Hall conductance becomes 
topological invariant and is quantized exactly\cite{QHE}\cite{QHE2}. 
In this paper, we consider a parity violating system 
in which gauge invariance is 
spontaneously broken. 
In such a system, these assumptions are not realized, so  
it is interesting and important to find out a structure of Chern-Simons term 
and its dynamical effects. 

 In Lorentz invariant system, 
Chern-Simons term can only exist in 2+1 dimension. But in non-relativistic 
systems, Chern-Simons term can also exist in 3+1 dimension, as well, in a form 
that is embedded in 2+1 dimension of 3+1 dimensional spacetime. 
~\\

 As a system in which these situations are simultaneously realized, 
we study rotating superfluid $^{3}$He-A \cite{and} \cite{He}. 
Although $^{3}$He atom is neutral, we can introduce 
external ``U(1) gauge field'' in order to consider 
rotation and other motion of  $^{3}$He-A system, and 
can introduce also gauge symmetry approximately with this field. 
This symmetry is spontaneously broken, however, 
by Cooper pair condensation. These pairs 
have orbital angular momentum 1 along some direction in $^{3}$He-A, so 
parity is also violated spontaneously. Furthermore, 
this system is non-relativistic, of course. 
It is a purpose of the present paper to study Chern-Simons term in this 
system and its physical implications.

Our paper is organized in the following manner. In section 2, 
we derive an action for rotating $^{3}$He-A. In section 3, we 
discuss about Chern-Simons term in 2+1 dimensional rotating $^{3}$He-A. 
Usually, Chern-Simons term is induced in an effective action for 
external gauge field from Fermion 1-loop diagram. 
But in the present case, Chern-Simons term is induced 
after perturbative calculation of Fermion and Goldstone mode in the 
lowest order, and its coefficient depends on an infra-red cut off of 
Goldstome mode. 
In section 4, we consider 3+1 dimensional case.

Owing to induced Chern-Simons term, there is a kind of ``Hall current'' in 
superfluid $^{3}$He-A. Furthermore, a problem about the orbital  
angular momentum of $^{3}$He-A system, so-called ``angular momentum paradox''
is resolved by Chern-Simons system. 
In section 5, we show that Hall current occurs at an edge of 
cylindrical superfluid $^{3}$He-A and this current gives an 
angular momentum. Summary are given in section 6.

\section{The action for superfluid $^{3}$He-A in rotating system} 

 We derive the effective action of 2+1 and 3+1 dimensional 
superfluid $^{3}$He-A in rotating system in this section. 
Although $^{3}$He atom is neutral, as we shall see below, 
we are able to introduce ``U(1) gauge field $A_{\mu}$'' in $^{3}$He action 
in rotating system and to study the system's properties with it. 

Let $\psi_{\alpha}(x)$ stand for $^{3}$He atom with spin $\alpha$. 
The original action for homogeneous $^{3}$He in the rest system is 
(D as the dimension of the system.)
\begin{eqnarray}
S_{{\rm{org.}}}=&\int d^{D}x 
\psi_{\alpha}^{\dagger}(x)
\{i \partial_{0}-(\frac{\vec{p}^{2}}{2m} - \epsilon_{{\rm F}} )\}\psi_{\alpha}(x) &\nonumber\\
&-\frac{1}{2}\int d^{D}x d^{D-1}x^{\prime} 
\psi_{\alpha}^{\dagger}(x)\psi_{\beta}^{\dagger}(x^{\prime})
V_{\alpha\beta;\gamma\delta}(\vec{x}-\vec{x^{\prime}}) 
\psi_{\gamma} (x^{\prime}) {\psi}_{\delta}(x), &\label{a1}
\end{eqnarray}\label{a1}\\
where $\epsilon_{{\rm F}}$ is Fermi energy and $V$ is instantaneous interaction
between neutral $^{3}$He atom. The transformation from rest frame coordinate $\{\vec{x}_{{\rm s}}\}$ to moving frame coordinate $\{\vec{x}(t)\}$ fixed to bulk $^{3}$He which is rotating around 3rd-axis (z-axis) with angular velocity $\Omega$ is 
\begin{equation}
\vec{x}_{{\rm s}} \rightarrow \vec{x}(t)=R_{D-1}(t) \vec{x}_{{\rm s}} ;
R_{2}(t)=\left( \begin{array}{cc}
           {\rm cos} \Omega t & - {\rm sin} \Omega t \\
           {\rm sin} \Omega t & {\rm cos} \Omega t  \end{array} \right),
R_{3}(t)=\left( \begin{array}{ccc}
           {\rm cos} \Omega t & - {\rm sin} \Omega t & 0 \\
           {\rm sin} \Omega t & {\rm cos} \Omega t & 0 \\
           0 & 0 & 1 \end{array} \right),
\end{equation}
and time derivative $\psi$ is transformed as
\begin{equation}
i \partial_{0} \psi \rightarrow  i \partial_{0} \psi + i \frac{d \vec{x}(t)}{d t} \cdot \vec{\nabla} \psi ; \frac{d \vec{x}(t)}{d t}=\vec{\Omega} \times \vec{x}.
\end{equation}
We define external ``vector potential'' as $\vec{A}=m \vec{\Omega} \times \vec{x}$, and external ``scalar potential $A_{0}$'' as difference of the chemical potential from its homogeneous value $\epsilon_{{\rm F}}$, then the rotating $^{3}$He action is written as 
\begin{eqnarray}
S_{{\rm rot.}}[\psi , \psi^{\dagger} , A_{\mu}]
=&\int d^{D}x \psi^{\dagger}(x)\{i \partial_{0}+A_{0}-(\frac{(\vec{p}+\vec{A})^{2}}{2m} - \epsilon_{{\rm F}} )+\frac{\vec{A}^{2}}{2m}\}\psi(x) &\nonumber\\
&-\frac{1}{2}\int d^{D}xd^{D-1}x^{\prime}{\psi}^{\dagger}(x){\psi}^{\dagger}(x^{\prime})V(\vec{x}-\vec{x^{\prime}}) {\psi} (x^{\prime}) {\psi}(x). &\label{a2}
\end{eqnarray}
Except $\psi^{\dagger} \frac{\vec{A}^{2}}{2m} \psi$ term, the action is invariant under a gauge transformation $A_{\mu} \rightarrow A_{\mu} + \partial_{\mu} \xi, \psi \rightarrow e^{i \xi} \psi$. 
The generating functional is defined by using path integral formalism as 
\begin{equation}
Z[A_{\mu}]= \int {\cal{D}}\psi^{\dagger} {\cal{D}}\psi e^{i S_{{\rm rot.}}[\psi ,\psi^{\dagger},A_{\mu}]}.\label{a3}
\end{equation}

We introduce a pair field $\Psi$ which specifies that $^{3}$He is in the superfluid state and rewrite Eq.(\ref{a3}). This can be done by introducing auxilially field and by transforming field variables in the same way as the Stratonovich-Hubbard transformation. Let us define a path integral of a Gaussian integral form, N, as
\begin{equation}
N=\int {\cal{D}}\Psi^{\dagger} {\cal{D}}\Psi e^{i \Delta S},\label{a4}
\end{equation}
\begin{eqnarray}
&\Delta S 
=\frac{1}{2} \int d^{D}x d^{D-1}x^{\prime} (\Psi_{\beta\alpha}^{\dagger}(x^{\prime},x)-\psi_{\alpha}^{\dagger}(x)\psi_{\beta}^{\dagger}(x^{\prime}))V_{\alpha\beta;\gamma\delta}(\vec{x}-\vec{x^{\prime}})&\nonumber\\
&\times (\Psi_{\gamma\delta}(x^{\prime},x)-\psi_{\gamma}(x^{\prime})\psi_{\delta}(x)),
&\nonumber
\end{eqnarray}
and insert Eq.(\ref{a4}) into Eq.(\ref{a3}). $Z[A_{\mu}]$ can be written as
\begin{equation}
Z[A_{\mu}]=\frac{1}{N}\int {\cal{D}}\psi^{\dagger}{\cal{D}}\psi{\cal{D}}\Psi^{\dagger}
{\cal{D}}\Psi e^{i S},\label{a5}
\end{equation}

\begin{eqnarray}
S = S_{{\rm rot.}}+\Delta S =
&\int d^{D}x 
\psi^{\dagger}(x)
\{i \partial_{0}+A_{0}-(\frac{(\vec{p}+\vec{A})^{2}}{2m} - \epsilon_{{\rm F}} )\}\psi(x)&\nonumber\\
&-\frac{1}{2}\int d^{D}x d^{D-1}x^{\prime}
\psi^{\dagger}(x)\psi^{\dagger}(x^{\prime})
V(\vec{x}-\vec{x^{\prime}})
\Psi(x^{\prime},x)&\nonumber\\
&-\frac{1}{2}\int d^{D}x d^{D-1}x^{\prime}
\Psi^{\dagger}(x^{\prime},x)
V(\vec{x}-\vec{x^{\prime}})
\psi(x^{\prime})\psi(x)&\nonumber\\
&+\frac{1}{2}\int d^{D}x d^{D-1}x^{\prime}
\Psi^{\dagger}(x^{\prime},x)
V(\vec{x}-\vec{x^{\prime}})
\Psi(x^{\prime},x) .&\nonumber
\end{eqnarray}
We regard $\Psi^{0}(\vec{x},\vec{x}^{\prime})$ as a stationary value of $\Psi(x,x^{\prime})$. This value is determined by minimizing an effective potential, 
which is derived by integrating out Fermion field ($\psi,\psi^{\dagger}$) 
while $A_{\mu}$ and dynamical fluctuation of $\Psi$ are kept constant. $\Psi^{0}$ satisfies  
\begin{equation}
\Psi_{\alpha\beta}^{0}(\vec{x},\vec{x}^{\prime})
=<\psi_{\alpha}(\vec{x})\psi_{\beta}(\vec{x}^{\prime})>.
\end{equation}
We can see easily that U(1) symmetry is spontaneously broken, 
whenever $\Psi^{0}$ has non zero value. The mean field solutions have 
been given in \cite{and}.

 In integrating $\Psi$ and $\Psi^{\dagger}$ in Eq.(\ref{a5}), 
we only consider the most important variable in the low energy reagion, 
i.e., phase degrees of freedom 
around $\Psi^{0}$ (Goldstone mode), 
and neglect fluctuations of other degrees of freedom. 
So, we write $\Psi$ as a product of $\Psi^{0}$ part, SU(2) part, 
and U(1) part,
\begin{equation}
\Psi_{\alpha\beta}(x,x^{\prime})
=e^{-i \theta(x) - i \theta(x^{\prime})}
U_{\alpha\alpha^{\prime}}(x)U_{\beta\beta^{\prime}}(x^{\prime})
\Psi_{\alpha^{\prime}\beta^{\prime}}^{0}(\vec{x},\vec{x}^{\prime}),
\end{equation}
$$
U(x)=e^{-i \phi^{a}(x)\cdot\sigma^{a}},
$$
where $\sigma^{a}$($a$=1,2,3) is Pauli matrices. We transform Fermion fields 
in the path integral (\ref{a5}) as
\begin{equation} 
\psi \rightarrow e^{- i \theta} U \psi , \psi^{\dagger} \rightarrow \psi^{\dagger} U^{\dagger} e^{i \theta}.
\end{equation}
This transformation does not change the path integral measure 
(i.e.${\cal{D}}\psi^{\prime\dagger} {\cal{D}}\psi^{\prime}
={\cal{D}}\psi^{\dagger} {\cal{D}}\psi$), 
so the action S becomes 
\begin{eqnarray}
S =
   &\int d^{D}x 
   \psi^{\dagger}(x)
   \{ i \partial_{0} + A_{0}+\partial_{0}\theta+iU^{\dagger}\partial_{0}U
    -(\frac{(\vec{p} + \vec{A} + \vec{\partial}\theta + 
      i U^{\dagger}\vec{\partial}U)^{2}}{2m} - \epsilon_{{\rm F}} )
   +\frac{\vec{A}^2}{2m}\}\psi(x)&\nonumber\\
   &-\frac{1}{2}\int d^{D}xd^{D-1}x^{\prime}
    ({\psi}^{\dagger}(x)\Delta (\vec{x},\vec{x^{\prime}})
    {\psi}^{\dagger}(x^{\prime})
    +{\psi}(x)\Delta ^{\dagger}(\vec{x},\vec{x^{\prime}}) 
    {\psi} (x^{\prime}))&\label{a6}\\
   &+\frac{1}{2}\int d^{D}x d^{D-1}x^{\prime}
   \Psi^{0 \dagger}(\vec{x}^{\prime},\vec{x})
   U^{\dagger}(x^{\prime}) U^{\dagger}(x)
   V(\vec{x}-\vec{x^{\prime}})U(x^{\prime})U(x)
   \Psi^{0}(\vec{x}^{\prime},\vec{x}) ,&\nonumber
\end{eqnarray}
where $\Delta, \Delta^{\dagger}$ are the gap functions, and they are determined as
\begin{equation}
\Delta_{\alpha\beta}(\vec{x}-\vec{x}^{\prime})
=V_{\alpha\beta;\gamma\delta}(\vec{x}-\vec{x}^{\prime}) 
\Psi_{\gamma\delta}^{0}(\vec{x}^{\prime},\vec{x}) 
\end{equation}
$$
\Delta_{\delta\gamma}^{\dagger}(\vec{x}^{\prime}-\vec{x})
=\Psi_{\beta\alpha}^{0\dagger}(\vec{x}^{\prime},\vec{x})
V_{\alpha\beta;\gamma\delta}(\vec{x}-\vec{x}^{\prime}),
$$
and in the action (\ref{a6}), contraction of spin suffices can be taken as 
$\psi_{\alpha}^{\dagger}\Delta_{\alpha\beta}\psi_{\beta}^{\dagger},
\psi_{\alpha}\Delta_{\alpha\beta}^{\dagger}\psi_{\beta}$, 
if the gap has parity odd structure.

In the superfluid $^{3}$He-A,  
the gap has angular momentum 1 along some direction. Hence, the gap is parity odd, and, parity symmetry is spontaneously broken in this system. We choose this angular momentum direction in 3rd-axis (z-axis) throughout our paper. In the weak coupling approximation, the gap has a form in the momentum space 
\cite{and} 
\begin{equation}
\Delta_{\alpha\beta}(\vec{k})=
\left\{ \begin{array}{l}
        i \Delta \frac{\sqrt{k_{1}^{2}+k_{2}^{2}}}{|\vec{k}|}
        e^{i \varphi} \delta_{\alpha\beta};
        (- \omega_{{\rm D}} < (\frac{\vec{k}^{2}}{2m}-\epsilon_{{\rm F}}) 
        < \omega_{{\rm D}})\\
        0;({\rm otherwise}), \end{array} \right.
\label{gap}\end{equation}
$$
{\rm tan}\varphi = k_{2}/k_{1},
$$
where $\omega_{{\rm D}}$ is Debey frequency, which satisfies the relation 
$|\Delta|<<\omega_{{\rm D}}<<\mu$.

Finally, the generating functional is written by,
\begin{equation}
Z[A_{\mu}]=\frac{1}{N}\int {\cal{D}}\theta{\cal{D}}\phi
{\cal{D}}\psi^{\dagger}{\cal{D}}\psi  e^{i S}.\label{a10}
\end{equation}

\section{Chern-Simons term in 2+1 dimensional superfluid $^{3}$He-A}

In this section, we calculate the right hand side of Eq.(\ref{a10}) 
in 2+1 dimensional case, and show 
how Chern-Simons term is induced. First, we integrate out Fermion fields 
in Eq.(\ref{a10}) and obtain $Z^{\rm (f)}$;
\begin{equation}
Z^{({\rm f})}[A_{\mu},\partial_{\mu} \theta,U^{\dagger} \partial_{\mu} U] 
= \int {\cal{D}}\psi^{\dagger} {\cal{D}}\psi e^{i S} 
= 
e^{iS_{{\rm eff.}}^{({\rm f})}
   [A_{\mu},\partial_{\mu} \theta,U^{\dagger} \partial_{\mu} U]}
.\label{a0}\end{equation}
To carry out this, we divide the action ${S}$ in two parts $S_{0}$, and 
$S_{\rm int.}$ as
\begin{equation}
S=S_{0} + S_{{\rm int.}}.
\end{equation}
$S_{0}$ is written as
\begin{equation}
S_{0}=\int d^{3}x d^{3}y \frac{1}{2} (\psi^{\dagger}(x) \psi(x))
\left( \begin{array}{cc} 
      iG(x-y) & F(x-y) \\
      -F^{\dagger}(x-y) & -iG(y-x) \end{array} \right)^{-1}
\left( \begin{array}{c}     
       \psi(y) \\
       \psi^{\dagger}(y) \end{array} \right),\label{s0}
\end{equation}
where
$$
\left\{ \begin{array}{c}
      iG(x-y)=<T\psi(x)\psi^{\dagger}(y)> \\
      F(x-y)=<T\psi(x)\psi(y)> \\
      F^{\dagger}(x-y)=<T\psi^{\dagger}(y)\psi^{\dagger}(x)>, \end{array} \right.
$$
are Fermion propagators in superfluid state, and their forms in momentum space
 obtained by solving Gor'kov equation are
\begin{equation}
\left\{ \begin{array}{c}
     iG(k)=\frac{k_{0}+(\frac{\vec{k}^{2}}{2m}-\epsilon_{{\rm F}})}
                {k_{0}^{2}-E^{2}(\vec{k})+i \epsilon} \\
     F(k)=\frac{i \Delta(\vec{k})}{k_{0}^{2}-E^{2}(\vec{k})+i \epsilon}\\
     F^{\dagger}(k)=\frac{-i \Delta^{\dagger}(\vec{k})}
                         {k_{0}^{2}-E^{2}(\vec{k})+i \epsilon}, 
        \end{array} \right.\label{fpro}
\end{equation}
$E(\vec{k})$ is quasiparticle energy and has the form $E(\vec{k})=
\sqrt{(\frac{\vec{k}^{2}}{2m}-\epsilon_{{\rm F}})^{2}+|\Delta(\vec{k})|^{2}}$, 
$\Delta(\vec{k})$ has the form (\ref{gap}) in the superfluid A phase 
and $|\Delta(\vec{k})|^{2}=|\Delta|^{2}={\rm const.}$ in 2+1 dimensional case.
$S_{{\rm int.}}$ is written as
\begin{eqnarray}
&S_{{\rm int.}}=\int d^{3}x 
[j_{0}(A_{0}+\partial_{0}\theta)
-\vec{j}\cdot(\vec{A}+\vec{\partial}\theta)
+\psi^{\dagger}\frac{(\vec{A}+\vec{\partial}\theta)^{2}}{2m}\psi
+\psi^{\dagger}\frac{\vec{A}^{2}}{2m}\psi &\nonumber\\
&+J_{0}^{a}(i U^{\dagger}\partial_{0}U)^{a}
-\vec{J}^{a}\cdot(U^{\dagger}\vec{\partial}U)^{a}
+\psi^{\dagger}\frac{(i U^{\dagger}\vec{\partial}U)^{2}}{2m}\psi],&\label{sint}
\end{eqnarray}
where $(j_{0},\vec{j})$ and $(J_{0}^{a},\vec{J}^{a})$ is U(1) and SU(2) currents which have forms
\begin{eqnarray}
&j_{0}=\psi^{\dagger}\psi,~~
 \vec{j}=-\frac{i}{2m}[(\vec{\partial}\psi^{\dagger})\psi 
                        -\psi^{\dagger}(\vec{\partial}\psi)]
 +\psi^{\dagger}\frac{(\vec{A}+\vec{\partial}\theta)}{m}\psi&\nonumber\\
&J_{0}^{a}=\psi^{\dagger} \sigma^{a} \psi,~~
 \vec{J}^{a}=-\frac{i}{2m}[(\vec{\partial}\psi^{\dagger})\sigma^{a}\psi 
                        -\psi^{\dagger}\sigma^{a}(\vec{\partial}\psi)]
 +\psi^{\dagger}\frac{(i U^{\dagger}\vec{\partial}U)^{a}}{m}\psi&.
\label{current}
\end{eqnarray}
So, $S_{{\rm eff.}}^{({\rm f})}$ should have the form
\begin{eqnarray}
&S_{{\rm eff.}}^{({\rm f})}=
\int d^{3}x d^{3}y 
[\frac{1}{2}(A_{\mu}+\partial_{\mu}\theta)_{x}\pi^{\mu\nu}(x-y)
(A_{\nu}+\partial_{\nu}\theta)_{y} -  
\frac{1}{2}\frac{\rho}{m}A_{i}(x)A^{i}(x)&\nonumber\\
&+\frac{1}{2}\partial_{\mu}\phi^{a}(x)\pi^{\mu\nu}_{ab}(x-y)\partial_{\nu}\phi^{b}(y)]+\cdot\cdot\cdot.&\label{a50}
\end{eqnarray}
We consider only lower dimensional terms in $A_{\mu}$ and 
$\partial_{\mu} \theta$, which comes from current 
2 point functions in the lowest order of 
perturbative calculation. It is possible to see that    
effects of spontaneous symmetry breaking are included in these terms. 
Scalar-Gauge interaction through current 2 point functions 
$A_{\mu}\pi^{\mu\nu}\partial_{\nu}\theta$  
do not appear if there exists U(1) gauge symmetry 
(for, Ward identity $\partial_{\mu} \pi^{\mu\nu}=0$).
But now, gauge symmetry is spontaneously broken so that this interaction
can exist. This interaction should give effects to the structure 
of Chern-Simons term in the spontaneously symmetry breaking case.

As far as current 2 point functions are concerned, SU(2) fields $\phi^{a}$ 
do not couple to $A_{\mu}$ and $\partial_{\mu}\theta$ 
because of spin conservation. 
So, we neglect field $\phi^{a}$ from now on.

In this action, $\frac{1}{2}\frac{\rho}{m}A_{i}(x)A^{i}(x)$ term
explicitly violates remained U(1) gauge invariance after spontaneously 
symmetry breaking; 
$A_{\mu} \rightarrow A_{\mu} + \partial_{\mu}\xi $, 
$\theta \rightarrow \theta - \xi $. 
Clearly, this violation comes from the existence of $\psi^{\dagger} \frac{\vec{A}^{2}}{2m} \psi$ term.

Now, one can derive current 2 point functions by the loop calculation 
about Fermion field. They are given in the momentum space as follows; 
\begin{equation}
\left\{ \begin{array}{l}
  \pi_{00}(p)=v^{2}+{\cal{O}}(p^{2})\\
  \pi_{0j}(p)
       =i\sigma_{xy}^{(0)} \varepsilon_{0ij} p_{i}+{\cal{O}}(p^{2}) \\
  \pi_{ij}(p)
       =-v^{2}c_{{\rm g}}^{2}\delta_{ij}+{\cal{O}}(p^{2})  \end{array} \right. 
\label{b8}
\end{equation}
with
\begin{equation}
\left\{ \begin{array}{l}
        \sigma_{xy}^{(0)}=\frac{1}{4\pi} \\
        v^{2}=N(0) \\
        v^{2}c_{{\rm g}}^{2}=\frac{\rho}{m} ,\end{array}
\right.
\label{b26}
\end{equation}
where $\rho$ is Fermion number density and $N(0)=\frac{m}{\pi}$ is density of state at
Fermi surface including spin degree of freedom in 2+1 dimension.

After scaling $\theta$ field as $\theta \rightarrow \theta/v$, 
$S_{{\rm eff.}}^{({\rm f})}$ becomes
\begin{eqnarray}
 &S_{{\rm eff.}}^{({\rm f})}=\int d^{3}x [\frac{v^{2}}{2}A_{0}^{2} 
  + \frac{\sigma_{xy}^{(0)}}{2} 
  \varepsilon_{0ij}(A_{0}\partial_{i}A_{j}+A_{i}\partial_{j}A_{0})
 +\frac{1}{2}\{(\partial_{0}\theta)^{2}-c_{g}^{2}(\vec{\partial} \theta)^{2}\}
 &\nonumber\\
 &+v A_{0}\partial_{0}\theta 
  -v c_{g}^{2} \vec{A} \cdot \vec{\partial} \theta 
  +\frac{\sigma_{xy}^{(0)}}{v} (\vec{\partial}\times\vec{A})
                                        (\partial_{0} \theta)]
  +\cdot\cdot\cdot.
 &\label{b5}
\end{eqnarray}

As we see in Eq.(\ref{a0}), $S_{\rm eff.}^{({\rm f})}$ is an effective action 
which is induced by dynamical effects of Fermion. 
We recognize $v$ and $c_{{\rm g}}$ as a decay constant and a velocity of sound 
of Goldstone mode.
The second and the third terms in the right hand side of Eq.(\ref{b5}) 
are parity violating and similar 
to the Chern-Simons term, but there is no 
$\varepsilon_{i0j}A_{i}\partial_{0}A_{j}$ term. 
This dose not contradict with gauge invariance. 
These terms comes from  
$(A_{0}+\partial_{0}\theta)\pi_{0i}
(A_{i}+\partial_{i}\theta)$ term, 
which is manifestly gauge invariant due to Goldstone mode.  
So, gauge invariance still remains without 
$\varepsilon_{i0j}A_{i}\partial_{0}A_{j}$ term 
owing to existence of Goldstone mode.

Because of the existence of this Chern-Simons like parity violating term, 
there is Hall current as a response to gradient of external scalar potential 
(chemical potential) $A_{0}$ which has the form
\begin{equation}
j^{i}=\sigma_{xy}^{(0)}\varepsilon^{0ij}\partial_{j}A_{0},\label{b15}
\end{equation}
and we can recognize $\sigma_{xy}^{(0)}$ as Hall conductance.
This result has been obtained by Volovik with slightly different 
manner from our present calculation\cite{Vol.}, and ours agree with 
his result.
\\

Next, we integrate out Goldstone mode and obtain the effective action of 
$A_{\mu}$. 
We see how Chern-Simons like parity
violating term in Eq.(\ref{b5}) has a correction. 
As we shall see in Eq.(\ref{b5}), 
$S_{{\rm eff.}}^{({\rm f})}$ has quadratic forms about Goldstone mode $\theta$ as far as bi-linear forms are concerned. Finally we have, 
\begin{eqnarray}
&e^{i S_{{\rm eff.}}[A_{\mu}]}=
\int {\cal{D}}\theta e^{i S_{{\rm eff.}}^{({\rm f})}}&\nonumber\\
&=det\{(\partial_{0}^{2} - c_{{\rm g}}^{2} \vec{\partial}^{2})^{-1}\} 
e^{i \int d^{3}x d^{3}y {\cal{J}}(x) D(x-y) {\cal{J}}(y)} 
e^{i \int d^{3}x \frac{\sigma_{xy}^{(0)}}{2} 
\varepsilon_{0ij}(A_{0}\partial_{i}A_{j}+A_{i}\partial_{j}A_{0})+\cdot\cdot\cdot}
&\label{b6}\\
\nonumber\\
&
=e^{i \int d^{3}x d^{3}y \frac{1}{2} A_{\mu}(x)\Pi^{\mu\nu}(x-y)A_{\nu}(y)
    +\cdot\cdot\cdot}.
&\label{b7}
\end{eqnarray}
$D(x-y)$ is Goldstone mode propagator
$$
D(x-y)=\int \frac{d^{3}p}{(2\pi)^{3}} \frac{1}{p_{0}^2 - c_{{\rm g}}^{2} |\vec{p}|^{2}},
$$
\begin{center}
\begin{minipage}[b]{5cm}
\epsfxsize=5cm \epsfbox{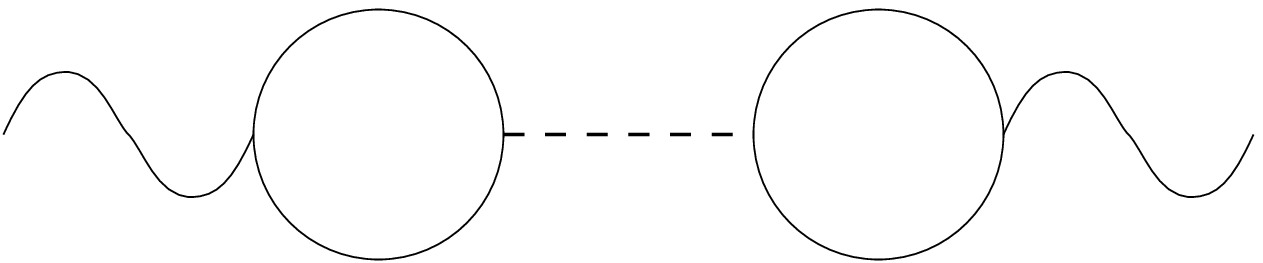}
\end{minipage}
~\\ 
Fig.1 Diagrams for Eq.(\ref{a80}). Dash line means Goldstone mode propagator 
$D(p)$.
Circles mean $\pi^{\mu\nu}(p)$.~~~~~~~~~~~~~~~~~~~~~~~~~~~~~~~~~~~~~~~~~~~~~~~

\end{center}
and ${\cal{J}}(x)$ is
$$
{\cal{J}}(x)=v \partial_{0} A_{0}
  - v c_{{\rm g}}^{2} \vec{\partial} \cdot \vec{A}
  + \frac{\sigma_{xy}^{(0)}}{v} (\vec{\partial}\times{\dot{\vec{A}}}).
$$
From Eqs.(\ref{b6}) and (\ref{b7}), we can 
evaluate $\Pi^{\mu\nu}$ in momentum space as

\begin{eqnarray}
&\Pi_{0i}(p)
=\pi_{0i}(p)+ \{ v^{2}c_{{\rm g}}^{2}p_{0}p_{i}
              +i\sigma_{xy}^{(0)}\varepsilon_{0ki}p_{0}^{2}p_{k} 
              +\cdot\cdot\cdot \}D(p)
&\nonumber\\
&
\Pi_{ij}(p)
=\pi_{ij}(p)
 -\{ v^{2}c_{{\rm g}}^{4}p_{i}p_{j}
 +ic_{{\rm g}}^{2}\sigma_{xy}^{(0)}p_{0}p_{k}(\varepsilon_{0kj}p_{i}
                                         -\varepsilon_{0ki}p_{j})&\nonumber\\
&~~~~~~~~~~+\frac{\sigma_{xy}^{(0)2}}{v^{2}}
            \varepsilon_{0ki}\varepsilon_{0lj}p_{k}p_{l}p_{0}
            +\cdot\cdot\cdot \} D(p).\label{a80}
&
\end{eqnarray}
These terms are shown diagramatically in Fig.1.

Linear terms in momentum is important in the low energy region, and  
have anti-symmetric structure about space-time indecies from $\Pi^{\mu\nu}$. 
Their coefficients can be written as 
\begin{eqnarray}
&
\frac{1}{2!} \varepsilon_{0ij}\frac{\partial}{\partial p_{i}}\Pi_{0j}(p)|_{p=0}
=i\sigma_{xy}^{(0)} [1-\frac{p_{0}^{2}}{p_{0}^2 - c_{{\rm g}}^{2} |\vec{p}|^{2}}|_{p=0}]
&\nonumber\\
&
\frac{1}{2!} \varepsilon_{i0j}\frac{\partial}{\partial p_{0}}
\Pi_{ij}(p)|_{p=0}=i\sigma_{xy}^{(0)}
  \frac{c_{{\rm g}}^{2}|\vec{p}|^{2}}{p_{0}^2 - c_{{\rm g}}^{2} |\vec{p}|^{2}}|_{p=0}
&\label{b11}
\end{eqnarray}
Here, some complications arise. 
The values $\frac{p_{0}^{2}}{p_{0}^2 - c_{{\rm g}}^{2} |\vec{p}|^{2}}|_{p=0},\frac{c_{{\rm g}}^{2}|\vec{p}|^{2}}{p_{0}^2 - c_{{\rm g}}^{2} |\vec{p}|^{2}}|_{p=0}$ in the right-hand side depend on how the limit is taken in the small momentum region. To see this clearly, we define $\epsilon$ as a  relative ratio 
between $p_{0}$ and $|\vec{p}|$ as follows;
\begin{equation}
\epsilon p_{0}=c_{{\rm g}}|\vec{p}|.\label{b12}
\end{equation}

Let us substitute Eq.(\ref{b12}) into Eq.(\ref{b11}), and then, 
we have $\Pi^{\mu\nu}(p)$
\begin{equation}
\Pi^{\mu\nu}(p)=\frac{\sigma_{xy}^{(0)}}{2}
                \frac{\epsilon^{2}}{\epsilon^{2}-1}
                \varepsilon^{\mu\rho\nu}ip_{\rho}+\cdot\cdot\cdot.
\end{equation}
Hence, $S_{{\rm eff.}}[A_{\mu}]$ in Eq.(\ref{b6}) has a term written as
\begin{equation}
S_{{\rm eff.}}^{\rm (C-S)}[A_{\mu}]=\int d^{3}x \frac{\sigma_{xy}^{(0)}}{2}
                  \frac{\epsilon^{2}}{\epsilon^{2}-1}
                  \varepsilon^{\mu\nu\rho}A_{\mu} \partial_{\nu} A_{\rho}
\end{equation}

So, we can see that ``Chern-Simons like'' term in 
$S_{{\rm eff.}}^{({\rm f})}$ of Eq.(\ref{b5}), has a correction from 
Goldstone mode and becomes{\it ``totally anti-symmetric Chern-Simons term''}. 
Hall current derived 
from this term has the form
\begin{equation}
j^{i}=\sigma_{{\rm He}}
        \varepsilon^{i0j}F_{0j}~~;~~
        \sigma_{{\rm He}}=
        \sigma_{xy}^{(0)}\frac{\epsilon^{2}}{\epsilon^{2}-1}.\label{b16}
\end{equation}

Concerning a behavior of Hall conductance,  
we see a significant difference between 
Quantum Hall effect and our case.
In Quantum Hall effect gauge invariance is strictly preserved, hence 
Hall conductance becomes topologically invariant, 
and is quantized exactly. The value becomes integer multiple of 
the fundamental constant that is proportional to fine structure constant.
But in the present case, gauge invariance is spontaneously broken and 
Hall conductance $\sigma_{{\rm He}}$ depends 
on an infra-red cut off. The low energy fluctuation due to Goldstone 
mode gives this effect. Consequently, the value in Eq.(\ref{b16}) 
is not strictly constant and depends on the boundary condition 
in temporal direction and spatial direction. In finite systems in 
spatial direction, the momentum cut off $|\vec{p}|$ is inversely 
proportional to the spatial size, but the energy cut off $p_{0}$ can be 
arbitrary small. Hence, $\epsilon$ should be infinite in this case, 
and $\sigma_{\rm He}$ in Eq.(\ref{b16}) agrees with $\sigma_{xy}^{(0)}$. 
In infinite systems, on the other hand, the momentum cut off $|\vec{p}|$ 
can be arbitrary small. If the energy cut off $p_{0}$ is small but finite, 
then $\epsilon$ vanishes, and $\sigma_{\rm He}$ in Eq.(\ref{b16}) vanishes 
in this case. The parameter $\epsilon$ can be of other finite value depending 
on the boundary conditions. Especially, if the small momentum limit is 
taken while the velocity is fixed with that of Goldstone mode, then 
$\epsilon=1$ ,and $\sigma_{{\rm He}}$ is divergent. 
In this special boundary condition $p_{0}=c_{{\rm g}}|\vec{p}|$, 
Goldstone mode causes resonance effect because 
this relation is the same as Goldstone mode's dispersion relation
and $\sigma_{{\rm He}}$ is enhanced. Varying $\epsilon$ around 1, 
Hall current changes its direction.

\section{Chern-Simons term in 3+1 dimensional Superfluid $^{3}$He-A}

In this section, We study 3+1 dimensional system, and derive Chern-Simons 
term in 3+1 dimensional rotating $^{3}$He-A.  
These systems are non-relativistic, so, Chern-Simons term embedded in 
2+1 dimension can exist. We show that this term is induced 
in these systems, actually. 
 
The calculation is almost the same as 2+1 dimensional case. 
First, we integrate out Fermion field as 
\begin{equation}
Z^{({\rm f})}[A_{\mu},\partial_{\mu} \theta,U^{\dagger} \partial_{\mu} U] 
= \int {\cal{D}}\psi^{\dagger} {\cal{D}}\psi e^{i S} 
= 
e^{iS_{{\rm eff.}}^{({\rm f})}[A_{\mu},\partial_{\mu} \theta,U^{\dagger} 
\partial_{\mu} U]}
,\end{equation} 
with the action of Eq.(\ref{a6}) in 3+1 dimension by perturbation expansion 
method.
Fermion propagators have the same forms as Eq.(\ref{fpro}), but now, absolute 
value of the gap has the following momentum dependence;
\begin{equation} 
|\Delta(\vec{k})|^{2}=|\Delta|^{2}(1-\frac{k_{z}^{2}}{|\vec{k}|^{2}})
\end{equation}
Clearly, this gap has two nodes in the z direction. Interactions 
are same as Eq.(\ref{sint}) with same currents (\ref{current}). 
Only difference is in the volume element $d^{3}x \rightarrow d^{4}x$ .
So, $S_{{\rm eff.}}^{({\rm f})}$ should be written as (c.f.Eq.(\ref{a50}))
\begin{eqnarray}
&S_{{\rm eff.}}^{({\rm f})}=
\int d^{4}x d^{4}y 
[\frac{1}{2}(A_{\mu}+\partial_{\mu}\theta)_{x}\pi^{\mu\nu}(x-y)
(A_{\nu}+\partial_{\nu}\theta)_{y} -   
\frac{1}{2}\frac{\rho}{m}A_{i}(x)A^{i}(x)&\nonumber\\
&+\frac{1}{2}\partial_{\mu}\phi^{a}(x)\pi^{\mu\nu}_{ab}(x-y)\partial_{\nu}\phi^{b}(y)]+\cdot\cdot\cdot.&\label{c15}
\end{eqnarray}
We neglect now SU(2) goldstone mode $\phi$ from the same reason as 2+1 
dimensional case.
Namely $\phi$ does not couple to $A_{\mu}$ and $\theta$ 
as far as current 2 point functions are concerned
because of spin conservation.

We calculate current 2 point functions perturbatively, 
and we have the current correlation functions in momentum space 
as 
\begin{equation}
\left\{ \begin{array}{l}
        \pi_{00}(p)=v^{2}+{\cal{O}}(p^{2})\\
        \pi_{0j}(p)
         =i\sigma_{xy\hat{z}}^{(0)} \varepsilon_{0ij\hat{3}} p_{i}
            +{\cal{O}}(p^{2})~~~;
        \varepsilon_{\mu\nu\rho\hat{3}}
         =\varepsilon_{\mu\nu\rho\sigma}(\hat{z})^{\sigma}\\
        \pi_{ij}(p)
       =-v^{2}c_{{\rm g}}^{2}\delta_{ij}+{\cal{O}}(p^{2})  \end{array} \right. 
\label{c10}
\end{equation}
with
\begin{equation}
\left\{ \begin{array}{l}
        \sigma_{xy\hat{z}}^{(0)}=\frac{N(0)}{4m} \\
        v^{2}=N(0) \\
        v^{2}c_{{\rm g}}^{2}=\frac{\rho}{m}
        ,\end{array} \right.
\label{c17}
\end{equation}
where $N(0)=\frac{3\rho}{2\epsilon_{{\rm F}}}$ is a density of state at Fermi 
surface including spin degree of freedom in 3+1 dimension.
These results are ${\it{not}}$ the same with that 
in 2+1 dimensional case (\ref{b26}).

Let us substitute Eqs.(\ref{c10}) and (\ref{c17}) into Eq.(\ref{c15}), 
and rescale $\theta$ as $\theta \rightarrow \theta/v$, then  
$S_{{\rm eff.}}^{({\rm f})}$ is written as 
\begin{eqnarray}
&S_{{\rm eff.}}^{({\rm f})}=\int d^{4}x [\frac{v^{2}}{2}A_{0}^{2}+\frac{\sigma_{xy\hat{z}}^{(0)}}{2} 
 \varepsilon_{0ij\hat{3}}(A_{0}\partial_{i}A_{j}+A_{i}\partial_{j}A_{0})
 +\frac{1}{2}\{(\partial_{0}\theta)^{2}
                -c_{{\rm g}}^{2}(\vec{\partial}\theta)^{2}\}&
 \nonumber\\
&+v A_{0}\partial_{0}\theta 
 -v c_{{\rm g}}^{2} \vec{A} \cdot \vec{\partial} \theta 
 +\frac{\sigma_{xy\hat{z}}^{(0)}}{v} (\vec{\partial}\times\vec{A})_{z}
                                        (\partial_{0} \theta)]
  +\cdot\cdot\cdot.
 &\label{c20}
\end{eqnarray}

The second and the thirs terms in the right hand side of Eq.(\ref{c20}) are 
parity violating terms and resemble Chern-Simons term. From this we 
can derive ``Hall current'' as follows;
\begin{equation}
\vec{j}
=\sigma_{xy\hat{z}}^{(0)}(\vec{\partial}A_{0} \times \hat{\vec{z}})
\label{b89}
\end{equation}

In $^{3}$He-A case, $A_{0}$ is chemical potential, and its relation 
with $^{3}$He atom number density $\rho$ is $\rho=N(0)A_{0}$. 
So, the current of Eq.(\ref{b89}) becomes
\begin{equation}
\vec{j}
=\frac{1}{4m}(\vec{\partial}\rho \times \hat{\vec{z}}).\label{MaMu}
\end{equation}
This current agrees with that derived by Mermin and Muzikar\cite{Mer.Muz.}. 

~\\
~\\

Next, we integrate out Goldstone mode and obtain an effective action, 
\begin{equation}
e^{i S_{{\rm eff.}}[A_{\mu}]}=\int {\cal{D}} \theta 
                              e^{i S_{{\rm eff.}}^{({\rm f})}},
\end{equation}
with parameters written in Eq.(\ref{c20}). 
By integrating Goldstone mode, 
we obtain a term in $S_{\rm eff.}[A_{\mu}]$ written as,  
\begin{equation}
S_{{\rm eff.}}^{(\rm C-S)}[A_{\mu}]=
\int d^{4}x \frac{\sigma_{xy\hat{z}}^{(0)}}{2}
\frac{\epsilon^{2}}{\epsilon^{2}-1}
\varepsilon^{\mu\nu\rho\sigma}A_{\mu} \partial_{\nu} A_{\rho}\hat{z}_{\sigma}.
\end{equation}
So, we see that in 3+1 dimensional rotating $^{3}$He-A,  
Chern-Simons term which is embedded 
in 2+1 dimensional space is induced by integrating 
Fermion and Goldstone mode.  

Hall current derived from this action is 
\begin{equation}
   j^{i}=\sigma_{{\rm He}}
        \varepsilon^{i0j\hat{3}}F_{0j}~~;~~
        \sigma_{{\rm He}}=
        \sigma_{xy\hat{z}}^{(0)}\frac{\epsilon^{2}}{\epsilon^{2}-1}.\label{c16}
\end{equation}

In the same manner as the 2+1 dimensional system, 
this current has Hall conductance. The magnitude, however, depends on 
an infra-red cut off of Goldstone mode.

\section{Orbital angular momentum of superfluid $^{3}$He-A in a cylinder}

We discuss a physical implications of Chern-Simons term. It has been known 
that there is a problem about the orbital angular momentum 
of $^{3}$He-A system. Cooper pairs in $^{3}$He-A have 
orbital angular momentum $l_{z}=1$. 
So, in a $^{3}$He-A system at $T=0$ in which Cooper pairs 
condensate homogeneously, it is expected that 
{\it total} angular momentum in the system $L_{z}$ coincides 
with a total number of 
pairs $\frac{N}{2}$ in natural unit $\hbar=1$ 
($N$; total number of $^{3}$He atom). 
Angular momentum {\it density} ${\cal{L}}_{z}$ in an 
isotropic and homogeneous system have been calculated with 
field theoretical method, but the expected result  
$\frac{\rho}{2}$ could not be obtained. 
This problem is so-called ``the angular momentum paradox of $^{3}$He-A''
\cite{Kita1}.
 
 Recently, it is shown by Kita \cite{Kita2} that a current exists at the edge of  
$^{3}$He-A in a cylinder which has axial symmetry around z-axis and  
whose Cooper pairs homogeneously condensate, and this edge 
current contributes to 
$L_{z}$ with the expected value, $\frac{N}{2}$. In this section, 
we show that one can easily obtain this edge current 
from our Hall current (\ref{c16}) by putting cylindrical  
boundary condition and can obtain $L_{z}=\frac{N}{2}$. 
The value may be enhanced if one puts on a spetial time dependent boundary 
condition in the system. 

Contribution of Hall current (\ref{c16}) to ${\cal{L}}_{z}$ is written as 
follows; 
\begin{equation}
{\cal{L}}_{z}=(\vec{x} \times m \vec{j})_{z} \label{d1}
\end{equation}
A boundary condition for $^{3}$He-A in a cylinder 
with radius $a$ is 
\begin{equation}
\rho(\vec{x})=\rho \theta(a-r)~,~\theta(r)~;{\rm the~step~function.}
\end{equation}
This condition is equivalent in our discussion to put on the $A_{0}$ and 
$\vec{A}$ as 
\begin{eqnarray}
&A_{0}(x)=\frac{1}{N(0)}\rho(\vec{x})&\nonumber\\
&\vec{A}(x)=0.&
\end{eqnarray}
Because, this boundary condition has no 
time dependence, the parameter $\epsilon$ 
should be infinite. So, Hall current (\ref{c16}) reduces to the form 
(\ref{MaMu}), 
and Eq.(\ref{d1}) becomes 
\begin{equation}
{\cal{L}}_{z}=\frac{-1}{4}\vec{x}\cdot\vec{\partial}\rho(\vec{x})
=\frac{\rho}{4}r \delta(a-r).
\end{equation}
Hence, $L_{z}$ is 
\begin{eqnarray}
L_{z}=& \int_{0}^{1}dz\int_{0}^{2\pi}d\theta\int_{0}^{a}dr r 
       {\cal {L}}_{z}&\nonumber\\
      &=\frac{\rho}{2}\pi a^{2}=\frac{N}{2}&     
\end{eqnarray}

If time dependent boundary conditions could be prepared, $L_{z}$ becomes,  
\begin{equation}
L_{z}=\frac{N}{2}\frac{\epsilon^{2}}{\epsilon^{2}-1},
\end{equation}
i.e., as same as Hall conductance, ${\cal{L}}_{z}$ is affected by 
spacetime dependence of boundary conditions. It is interesting if this 
different value will be observed.

\section{Summary}
 We have considered 2+1 and 3+1 dimensional rotating superfluid $^{3}$He-A. 
In rotating $^{3}$He-A, we could introduce external U(1) gauge field 
in the system, and this approximate gauge invariance is 
spontaneously broken by Cooper pair condensation. 
We have shown that Abelian Chern-Simons term is 
induced in the effective action for 
external gauge field from the combined effects of Fermion and Goldstone mode. 
In quantum Hall effect, gauge invariance is strictly preserved 
and, Hall conductance which is a coefficient of 
Chern-Simons term becomes topological invariant 
and it is quantized exactly. But in the present case, 
we have shown that Hall conductance 
should depend on an infra-red cut off of Goldstone mode. 
This infra-red cut off have a relation with 
spacetime dependence of boundary condition and, 
Hall conductance is enhanced when 
Goldstone mode causes resonance. 

 We have shown that the total orbital angular momentum $L_{z}$ in 
cylindrical superfluid $^{3}$He-A is derived automatically 
from Chern-Simons term and, $L_{z}$ is proportional 
to Hall conductance. So, $L_{z}$ also depends on infra-red cut off 
of Goldstone mode, and varies its value with changing spacetime dependence 
of boundaly condition. If a system has no time dependent boundaly condition 
with homogeneous pair condensation, $L_{z}$ takes the 
value $\frac{N}{2}$ which is a total number of Cooper pairs. 
It is the expected value under the problem 
which is called ``the angular momentum paradox in $^{3}$He-A''. 
This value is changed, however, if we put on a time dependent 
boundary conditions.  

 These results are obtained by a perturbative calculation in the lowest order,
expressed in (Fig.1). In the system without spontaneous breaking of 
gauge symmetry, there is no correction to Hall conductance 
from higher order effects.  
In our case, gauge symmetry is spontaneously broken   
and there could be higher order corrections. 
It is our future probrem to study higher order corrections 
to Hall conductance from diagrams of Fig.2.     

\begin{center}
\begin{minipage}[b]{5cm}
\epsfxsize=5cm \epsfbox{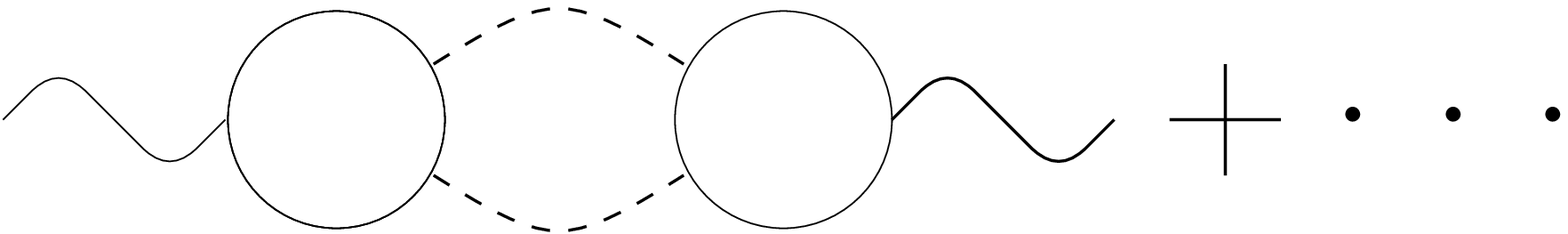}
\end{minipage}
Fig.2 Higher order correction terms. Circles mean 3 point functions which is 
induced by Fermion loop calculation. 
\end{center}

\begin{center}
{\bf ACKNOWLEDGMENT}
\end{center}

The authors are grateful to Professor T. Kita and 
Dr. N. Maeda for useful discussions.

This work was partially supported by the special Grant-in-Aid for Promotion 
of Education and Science in Hokkaido University provided by the Ministry 
of Education, Science, Sports, and Culture, the Grant-in-Aid for Scientific 
Research (07640522), the Grant-in-Aid for Scientific Research on Priority area
(Physics of CP violation), and the Grant-in-Aid for International Science 
Research (Joint Research 07044048) from the Ministry of Education, Science, 
Sports and Culture, Japan.   


\end{document}